\documentstyle[twoside,fleqn,npb,epsfig,amssymb,euscript]{article}
\def\Order#1{${\cal O}(#1)$}

\def\Ordex#1{${\cal O}(#1)_{exp}$}

\def\bbeta{\bar{\beta}}
\def\hbeta{\hat{\beta}}

\newcommand{\KK}{${\cal KK}$}

\newcommand{\Meu}{\EuScript{M}}

\newcommand{\Mmf}{\mathfrak{M}}

\def\st{\hbox{}} 

\newcommand{\AmS}{{\protect\the\textfont2
  A\kern-.1667em\lower.5ex\hbox{M}\kern-.125emS}}
\newcommand{\sfac}{\mathfrak{s}}
\title{Precision calculation for $e^+e^-\rightarrow 2f$:
the ${\cal KK}$ MC project\thanks{Work partly supported 
by EU contract HPRN-CT-2000-00149, by NATO Grant PST.CLG.977751, and by
Polish Government grants 5P03B09320 and 2P03B00122 and by US Department of Energy Contract  DE-FG05-91ER40627.}}
\author{B.F.L. Ward\address{Department of Physics and Astronomy, University of Tennessee, \\
Knoxville, Tennessee 37996-1200, USA}
     and   
    {S. Jadach$^{\rm b}$ and Z. Was}\address{Henryk Niewodniczanski Institute of Nuclear Physics, \\
        ul. Radzikowskiego 152, 31-342 Cracow, Poland}}
\begin{document}
\begin{abstract}
We present the current status of the coherent exclusive (CEEX)
realization of the YFS theory for the processes in $e^+e^-\rightarrow 2f$
via the ${\cal KK}$ MC. We give a brief summary of the
CEEX theory in comparison to the older (EEX) exclusive exponentiation theory
and illustrate recent theoretical results relevant to the LEP2 and LC
physics programs.
\centerline{UTHEP-02-0901}
\centerline{Sept, 2002}
\vspace{1pc} 
\end{abstract}

\maketitle

\section{Introduction}
Our aims in this discussion are to summarize briefly on
the main features of YFS/CEEX exponentiation~\cite{ceex1} in QED and
to present examples of recent theoretical results~\cite{recent1,recent2} 
relevant for the 
LEP/LC physics programs.

In the next section, we review the older EEX exclusive realization 
and summarize the new CEEX exclusive realization of the
YFS~\cite{yfs} theory in QED. In this way we illustrate the latter's 
advantages over the former, which is also very successful. 
We also stress the key common aspects of our MC implementations of 
the two approaches to exponentiation,
such as the exact treatment of phase space in both cases, 
the strict realization of the factorization theorem, etc.
In Sect. 3, we illustrate recent improvements in the ${\cal KK}$ MC realization
of CEEX for the $\nu\bar\nu$ channel. In Sect. 4 we illustrate
recent exact results on the single hard bremsstrahlung in 2f processes which
quantify the size of the missing sub-leading ${\cal O}(\alpha^2)$L
terms in the ${\cal KK}$ MC. Sect. 5 contains our summary. 

\section{Standard Model calculations for LEP with YFS exponentiation}
There are currently many successful applications~\cite{eex} of the YFS
theory of exponentiation for LEP and LC physics: (1), for 
$e^+e^- \to f\bar{f} +n\gamma$, $f=\tau,\mu,d,u,s,c$ there are
YFS1 (1987-1989) \Ordex{\alpha^1} ISR,
YFS2$\in$KORALZ (1989-1990), \Ordex{\alpha^1+h.o.LL} ISR,
YFS3$\in$KORALZ (1990-1998), \Ordex{\alpha^1+h.o.LL} ISR+FSR, and
\KK\ MC (98-02) \Ordex{\alpha^2+h.o.LL} ISR+FSR+IFI with 
$d\sigma/\sigma = 0.2\%$; (2), for 
$e^+e^- \to e^+e^-+n\gamma$ for $\theta < 6^\circ$ there are 
BHLUMI 1.x, (1987-1990), \Ordex{\alpha^1} and
BHLUMI 2.x,4.x, (1990-1996), \Ordex{\alpha^1+h.o.LL} 
with $d\sigma/\sigma = 0.07\%$; (3), for 
$e^+e^- \to e^+e^-+n\gamma$ for $\theta > 6^\circ$ there
is BHWIDE (1994-1998), \Ordex{\alpha^1+h.o.LL} with
$d\sigma/\sigma = 0.2(0.5)\%$ at the Z peak ( just off the Z peak ); (4),
for $e^+e^- \to W^+W^-+n\gamma$, $W^\pm \to f\bar{f}$ there is
KORALW (1994-2001); and, (5), for 
$e^+e^- \to W^+W^-+n\gamma$, $W^\pm \to f\bar{f}$ there is 
YFSWW3 (1995-2001), YFS exponentiation + Leading Pole Approximation
with  $d\sigma/\sigma = 0.4\%$ at LEP2 energies above the WW threshold.
The typical MC realization we effect is in the form of 
``matrix element $\times$ exact phase space'' principle,
as we illustrate in the following diagram:
\vspace{-2mm}

\begin{minipage}{52mm}
{\epsfig{file=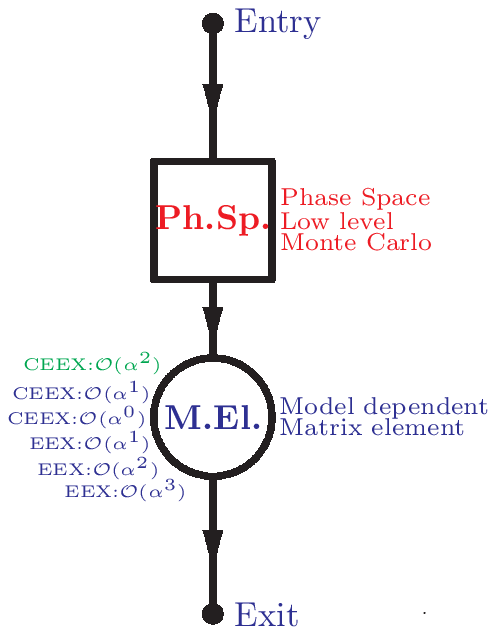,width=50mm}}
\end{minipage}
\vspace{4mm}
\begin{minipage}{52mm}
{\small
In practice it means:
\begin{itemize}
\item The universal exact Phase-space MC simulator is a separate module
      producing ``raw events'' (with importance sampling).
\item The library of several types of SM/QED matrix elements which provides
      the ``model weight'' is another independent module ( the \KK MC example is shown ).
\item Tau decays and hadronization come afterwards of course.
\end{itemize}}
\end{minipage}

The main steps in YFS exponentiation are the reorganization of the perturbative complete \Order{\alpha^\infty} series such that IR-finite $\bbeta$ components are isolated (factorization theorem) and 
the truncation of the IR-finite $\bbeta$s to finite \Order{\alpha^n}
with the attendant calculation of them from Feynman diagrams recursively.
We illustrate here the respective factorization for 
overlapping IR divergences for 
the 2$\gamma$ case -- $R_{12}\in R_1$ and $R_{12}\in R_2$ as they are
shown in the following picture:
\vspace{-4mm}
\begin{center}
  \fbox{\epsfig{file=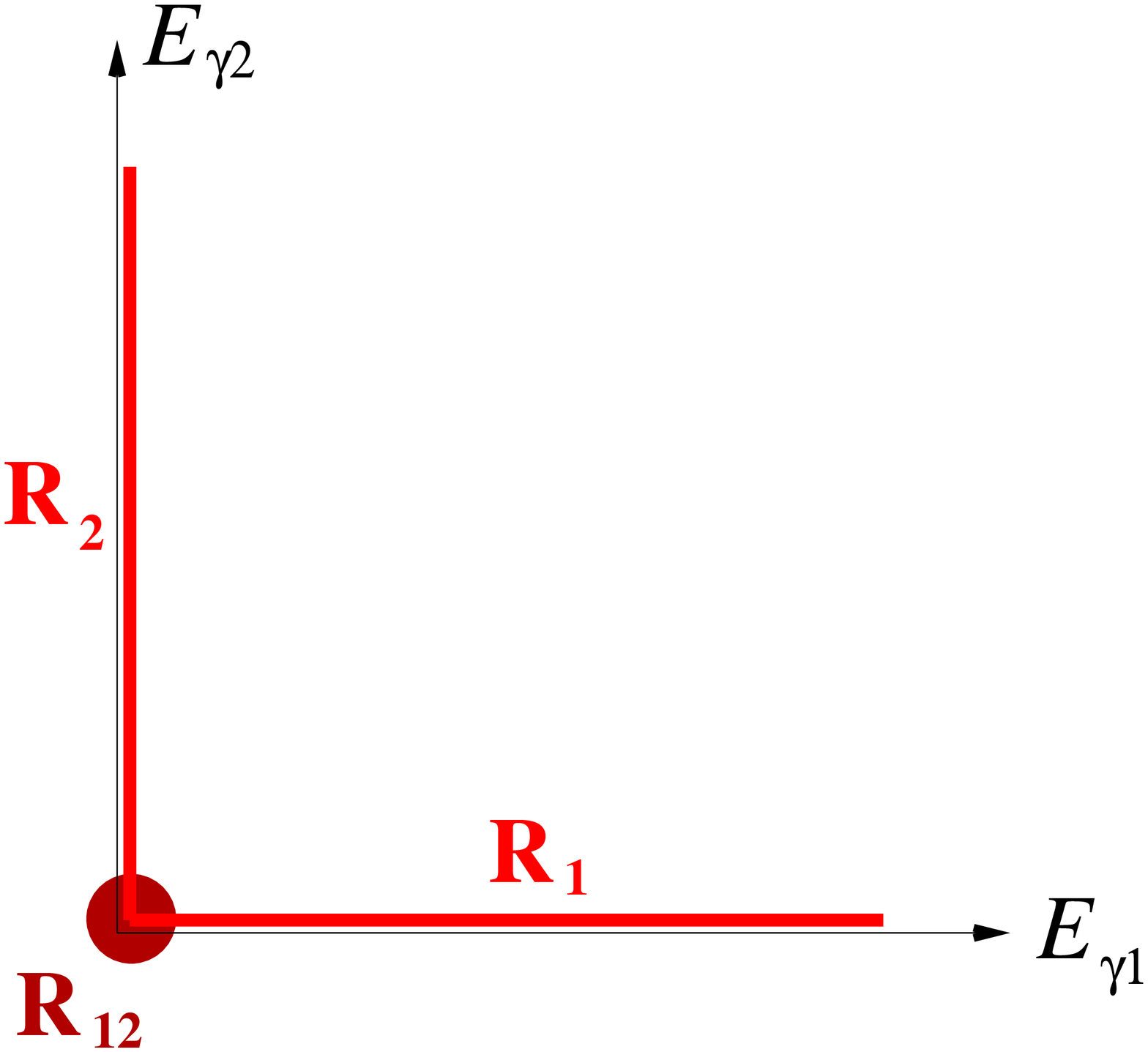,width=45mm,angle=270}}
\end{center}
\noindent
$D_0(p_{f_1},p_{f_2},p_{f_3},p_{f_4}) =
      \bbeta_0(p_{f_1},p_{f_2},p_{f_3},p_{f_4})
;\\ ~~~~~~ p_{f_1}+p_{f_2}=p_{f_3}+p_{f_4}$\\ 
$D_1(p_f;k_1) = 
        \bbeta_0(p_f) \tilde{S}(k_1) +\bbeta_1(p_f;k_1) 
;\\ ~~~~~~ p_{f_1}+p_{f_2}\neq p_{f_3}+p_{f_4}$\\
$D_2(k_1,k_2) =
        \bbeta_0 \tilde{S}(k_1)\tilde{S}(k_2) 
       +\bbeta_1(k_1)\tilde{S}(k_2)+\bbeta_1(k_2)\tilde{S}(k_1)
       +\bbeta_2(k_1,k_2)$.\\
\noindent Note:
$\bbeta_0$ and $\bbeta_1$ are used beyond their usual (Born and $1\gamma$) phase space. A kind of smooth ``extrapolation'' or ``projection'' is always necessary. We see that a recursive order-by-order calculation of the 
IR-finite $\bbeta$s to a given fixed \Order{\alpha^n} is possible: specifically,\\
$\bbeta_0(p_{f_1},p_{f_2},p_{f_3},p_{f_4})= D_0(p_{f_1},p_{f_2},p_{f_3},p_{f_4})$,\\
$\bbeta_1(p_f;k_1) = 
        D_1(p_f;k_1) -\bbeta_0(p_f) \tilde{S}(k_1) $,\\
$\bbeta_2(k_1,k_2) = D_2(k_1,k_2)
       -\bbeta_0 \tilde{S}(k_1)\tilde{S}(k_2) 
       -\bbeta_1(k_1)\tilde{S}(k_2)+\bbeta_1(k_2)\tilde{S}(k_1)$,
$\ldots$,
allow such a truncation.

In the classic EEX/YFS schematically the $\beta$'s are truncated to \Order{\alpha^1}, in the ISR example. For
{\scriptsize
    $e^-(p_1,\lambda_1)+e^+(p_2,\lambda_2)
    \to f(q_1,\lambda'_1)+\bar{f}(q_2,\lambda'_2)+\gamma(k_1,\sigma_1)+...+\gamma(k_n,\sigma_n)$ }, we have\\
{\small\begin{equation}
\sigma = \sum\limits_{n=0}^\infty \;\;
          \int\limits_{ m_\gamma} 
          d\Phi_{n+2}\; e^{Y(m_\gamma)} 
          D_n(q_1,q_2,k_1,...,k_n)
\end{equation}}
with
\begin{eqnarray}
D_0 = \bbeta_0,\qquad D_1(k_1) = \bbeta_0 \tilde{S}(k_1) +\bbeta_1(k_1),\nonumber\cr
D_2(k_1,k_2) = \bbeta_0 \tilde{S}(k_1)\tilde{S}(k_2)
                    +\bbeta_1(k_1)\tilde{S}(k_2)\nonumber\cr
+\bbeta_1(k_2)\tilde{S}(k_1)\qquad\qquad,\nonumber\cr
D_n(k_1,k_2...k_n)\;\;=\;\;\bbeta_0 \tilde{S}(k_1)\tilde{S}(k_2)...\tilde{S}(k_n)\nonumber\cr
    +\bbeta_1(k_1)\tilde{S}(k_2)\tilde{S}(k_3)...\tilde{S}(k_n)\qquad\qquad\nonumber\cr
    +\tilde{S}(k_1)\bbeta_1(k_2)\tilde{S}(k_3)...\tilde{S}(k_n)+...\qquad\nonumber\cr
    +\tilde{S}(k_1)\tilde{S}(k_2)\tilde{S}(k_3)...\bbeta_1(k_n).
\end{eqnarray}
The real soft factors are
{
    $\tilde{S}(k) = \sum\limits_\sigma |{\sfac}_\sigma(k)|^2 = |{\sfac}_+(k)|^2 +|{\sfac}_-(k)|^2
                  = -{\alpha\over 4\pi^2}\big({q_1\over kq_1}-{q_2\over kq_2} \big)^2 $}\\
{and the IR-finite building blocks are}
\\
$\bbeta_0= ( e^{-2\alpha\Re B_4} 
   \sum_\lambda |{\Meu}^{\rm Born+Virt.}_\lambda|^2 )\big|_{_{{\cal O}(\alpha^1)}}$, 
with $\lambda$ = fermion hel.,  $\sigma$ = photon hel., and
\\
$\bbeta_1(k)=  \sum\limits_{\lambda\sigma} |{\Meu}^{\rm 1-PHOT}_{\lambda\sigma}|^2 
                  -\sum\limits_{\sigma}  |{\sfac}_\sigma(k)|^2
                   \sum\limits_{\lambda} |{\Meu}^{\rm Born}_\lambda|^2$.
\\
{Everything is in terms of $\sum\limits_{spin} |...|^2 $ !}\quad
{ Distributions $<0$ are possible for hard $2\gamma$.}

The new CEEX replaces old the EEX, where both are derived from 
the YFS theory~\cite{yfs}: EEX, Exclusive EXponentiation, 
is very close to the original Yennie-Frautschi-Suura formulation; 
CEEX, Coherent EXclusive exponentiation, is an extension of the YFS
theory. Its coherence makes CEEX friendly to quantum coherence
among the Feynman diagrams, so that we have the complete
$|\sum\limits_{diagr.}^n {\Meu}_i\big|^2$ rather than the often incomplete
$\sum\limits_{i,j}^{n^2} {\Meu}_i {{\Meu}_j}^*$. This means 
be  get readily the proper treatment of
narrow resonances, $\gamma\oplus Z$~exchanges, $t\oplus s$~channels,
ISR$\oplus$FSR, angular ordering,~etc. Examples of the 
EEX formulation are KORALZ/YFS2, BHLUMI, BHWIDE, YFSWW, KoralW and  KORALZ;
the example of the CEEX formulation is \KK MC.\par

We illustrate CEEX schematically with the example of ISR \Order{\alpha^1}
for the process
    $e^-(p_1,\lambda_1)+e^+(p_2,\lambda_2)
    \to f(q_1,\lambda'_1)+\bar{f}(q_2,\lambda'_2)+\gamma(k_1,\sigma_1)+...+\gamma(k_n,\sigma_n)$. We have\\
{\small $\sigma = \sum\limits_{n=0}^\infty\;
          \int\limits_{m_\gamma} d\Phi_{n+2}\!\!\!\!
          \sum\limits_{\lambda,\sigma_1,...,\sigma_n}\!\!\!\!
          |e^{\alpha B(m_\gamma)}\\ 
 ~~~~~~~~~~~~~{\Meu}^{\lambda}_{n,\sigma_1,...,\sigma_n}(k_1,...,k_n)|^2 $,} 
\\
$ {\Meu}_{0}^{\lambda} = \hbeta_0^\lambda$,\quad $\lambda$=fermion helicities,\\
$ {\Meu}^\lambda_{1,\sigma_1}(k_1) 
              = \hbeta^\lambda_0 {\sfac}_{\sigma_1}(k_1) 
               +\hbeta^\lambda_{1,\sigma_1}(k_1)$,\\
$ {\Meu}^\lambda_{2,\sigma_1,\sigma_2}(k_1,k_2) 
              = \hbeta^\lambda_0 {\sfac}_{\sigma_1}(k_1) {\sfac}_{\sigma_2}(k_2)
               +\\
~~~~~~~~~\hbeta^\lambda_{1,\sigma_1}(k_1){\sfac}_{\sigma_2}(k_2)
               +\hbeta^\lambda_{1,\sigma_2}(k_2){\sfac}_{\sigma_1}(k_1) $,\\
{
$  {\Meu}^\lambda_{n,\sigma_1,...\sigma_n}(k_1,...k_n) = \hbeta^\lambda_0 {\sfac}_{\sigma_1}(k_1) {\sfac}_{\sigma_2}(k_2)\\~~~~~~~~...{\sfac}_{\sigma_n}(k_n)
     +\hbeta^\lambda_{1,\sigma_1}(k_1) {\sfac}_{\sigma_2}(k_2)...{\sfac}_{\sigma_n}(k_n)
$\\
$~~~~~~~
     +{\sfac}_{\sigma_1}(k_1) \hbeta^\lambda_{1,\sigma_2}(k_2)... {\sfac}_{\sigma_n}(k_n)+...
     +\\
~~~~~~~~{\sfac}_{\sigma_1}(k_1) {\sfac}_{\sigma_2}(k_2)...  \hbeta^\lambda_{1,\sigma_2}(k_2)
$.}

{The \Order{\alpha^1} IR-finite building blocks are:}\\
$\hbeta^\lambda_0 = \big(e^{-\alpha B_4} {\Meu}^{\rm Born+Virt.}_{\lambda}\big)\big|_{{\cal O}(\alpha^1)},$\\
$\hbeta^{\lambda}_{1,\sigma}(k)={\Meu}^\lambda_{1,\sigma}(k) - \hbeta^\lambda_0 {\sfac}_{\sigma}(k)$

{Everything is done in terms of ${\Meu}$-amplitudes!}\\
{Distributions are $\geq 0$ by construction!}\\
{In \KK MC the above is done up to \Order{\alpha^2} for ISR and FSR.}

 The full scale CEEX \Order{\alpha^r}, r=1,2, master formula 
for the polarized total cross section reads as follows:
{\small
 \begin{eqnarray}
~&\sigma^{(r)}\! = \!\!
  \sum_{n=0}^\infty {1\over n!}
  \int d\tau_{n} ( p_a\!+\!p_b ; p_c,p_d, k_1,\dots,k_n)\nonumber\\
 & e^{ 2\alpha\Re B_4 }\!
    \sum_{\sigma_i,\lambda,\bar{\lambda}}\;
    \sum_{i,j,l,m=0}^3
    \hat{\varepsilon}^i_a \hat{\varepsilon}^j_b
    \sigma^i_{\lambda_a \bar{\lambda}_a} \sigma^j_{\lambda_b \bar{\lambda}_b}\nonumber\\
   & \Mmf^{(r)}_n 
    \left(\st^{p}_{\lambda} \st^{k_1}_{\sigma_1} \st^{k_2}_{\sigma_2}
                                           \dots \st^{k_n}_{\sigma_n} \right)
    \Big[\Mmf^{(r)}_n\!\left(  \st^{p}_{\bar{\lambda}} \st^{k_1}_{\sigma_1} \st^{k_2}_{\sigma_2}\dots \st^{k_n}_{\sigma_n} \right)\Big]^\star\nonumber\\ 
& \sigma^l_{\bar{\lambda}_c \lambda_c } \sigma^m_{\bar{\lambda}_d \lambda_d } \hat{h}^l_c              \hat{h}^m_c.\nonumber\\
 \end{eqnarray}
}

{The respective CEEX amplitudes are}
{\small
\begin{eqnarray}
&\Mmf^{(1)}_n\left(\st^{p}_{\lambda} \st^{k_1}_{\sigma_1}
                              \dots \st^{k_n}_{\sigma_n}\right)\!
  =\sum\limits_{\wp\in{\cal P}}
   \prod\limits_{i=1}^n  {\sfac}_{[i]}^{\{\wp_i\}}
   \Bigg\{ \beta_0^{(1)}\big(\st^{p}_{\lambda};X_\wp\big) 
     \! \nonumber\\
&+\! { \sum\limits_{j=1}^n}
        {\beta^{(1)}_{1\{\wp_j\}}\big(\st^{p}_{\lambda}\st^{k_j}_{\sigma_j};X_{\wp}\big) 
             \over {\sfac}_{[j]}^{\{\wp_j\}} }
   \Bigg\}
\\
&\Mmf^{(2)}_n\left(\st^{p}_{\lambda} \st^{k_1}_{\sigma_1}
                              \dots \st^{k_n}_{\sigma_n}\right)
  =\!\sum\limits_{\wp\in{\cal P}}
   \prod\limits_{i=1}^n  {\sfac}_{[i]}^{\{\wp_i\}}\;\;\;\;\;\nonumber\\
  & ~~~~\times
   \Bigg\{ \beta_0^{(2)}\big(\st^{p}_{\lambda};X_\wp\big) 
   \!+\! { \sum\limits_{j=1}^n}
   {\beta^{(1)}_{2\{\wp_j\}}\big(\st^{p}_{\lambda}\st^{k_j}_{\sigma_j};X_{\wp}\big) 
             \over {\sfac}_{[j]}^{\{\wp_j\}} }
   \! \nonumber\\
&+\! { \sum\limits_{1\leq j<l\leq n}}
   {\beta^{(2)}_{2\{\wp_j,\wp_l\}}
    \big( \st^{p}_{\lambda}\st^{k_j}_{\sigma_j} \st^{k_l}_{\sigma_l}; X_{\wp}\big) 
         \over {\sfac}_{[j]}^{\{\wp_j\}}  {\sfac}_{[l]}^{\{\wp_l\}} }
   \Bigg\}.\nonumber\\
\end{eqnarray}
}
{For the details see ref.~\cite{ceex1}.}

\begin{figure}[htb]
\setlength{\unitlength}{1mm}
\vspace{9pt}
\framebox[55mm]{\rule[-21mm]{0mm}{43mm}}
\put(-50,-21){\makebox(0,0)[lb]{\epsfig{file=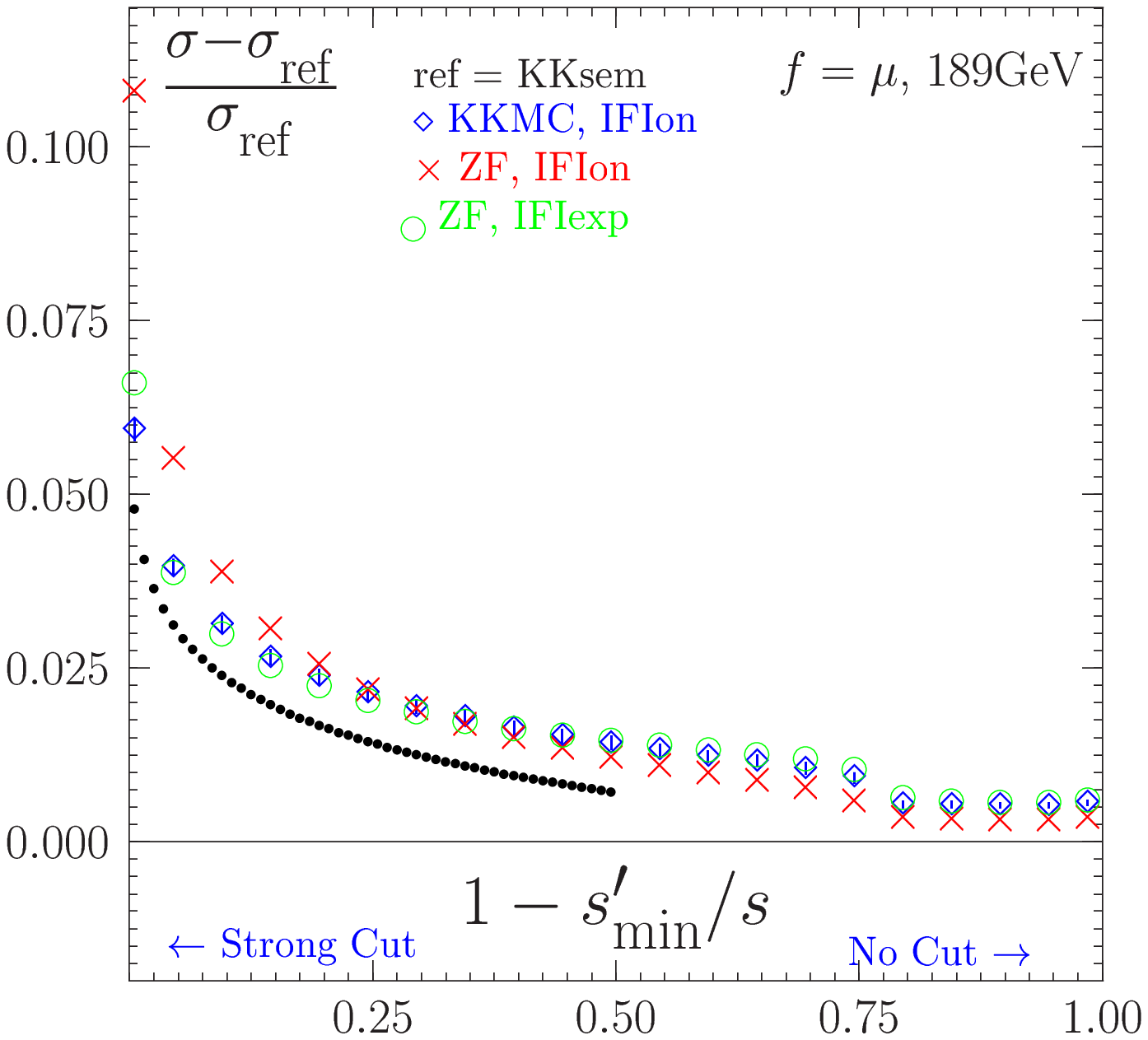,width=45mm}}}
\caption{Cross checks of \KK MC.}
\label{zfit}
\end{figure}

The precision tags of the \KK MC are determined by comparisons with
our own semi-analytical and independent MC results and by comparison
with the semi-analytical results of the program ZFITTER~\cite{zfitter}.
In Fig.~\ref{zfit} we illustrate such comparisons, which lead to
the \KK MC precision tag $d\sigma/\sigma = 0.2\%$ for example.
The ISR of ZFITTER is based on the \Order{\alpha^2}
result of ref.~\cite{berends},
while \KK MC is totally independent!
See ref.~\cite{ceex1} for a more complete discussion.

\section{Extension of CEEX in \KK MC to the $e^+e^-\to\nu\bar\nu\gamma$ process}
\begin{figure}[htb]
\setlength{\unitlength}{1mm}
\framebox[70mm]{\rule[-21mm]{0mm}{50mm}}
\put(-67,-21){\makebox(0,0)[lb]{\epsfig{file=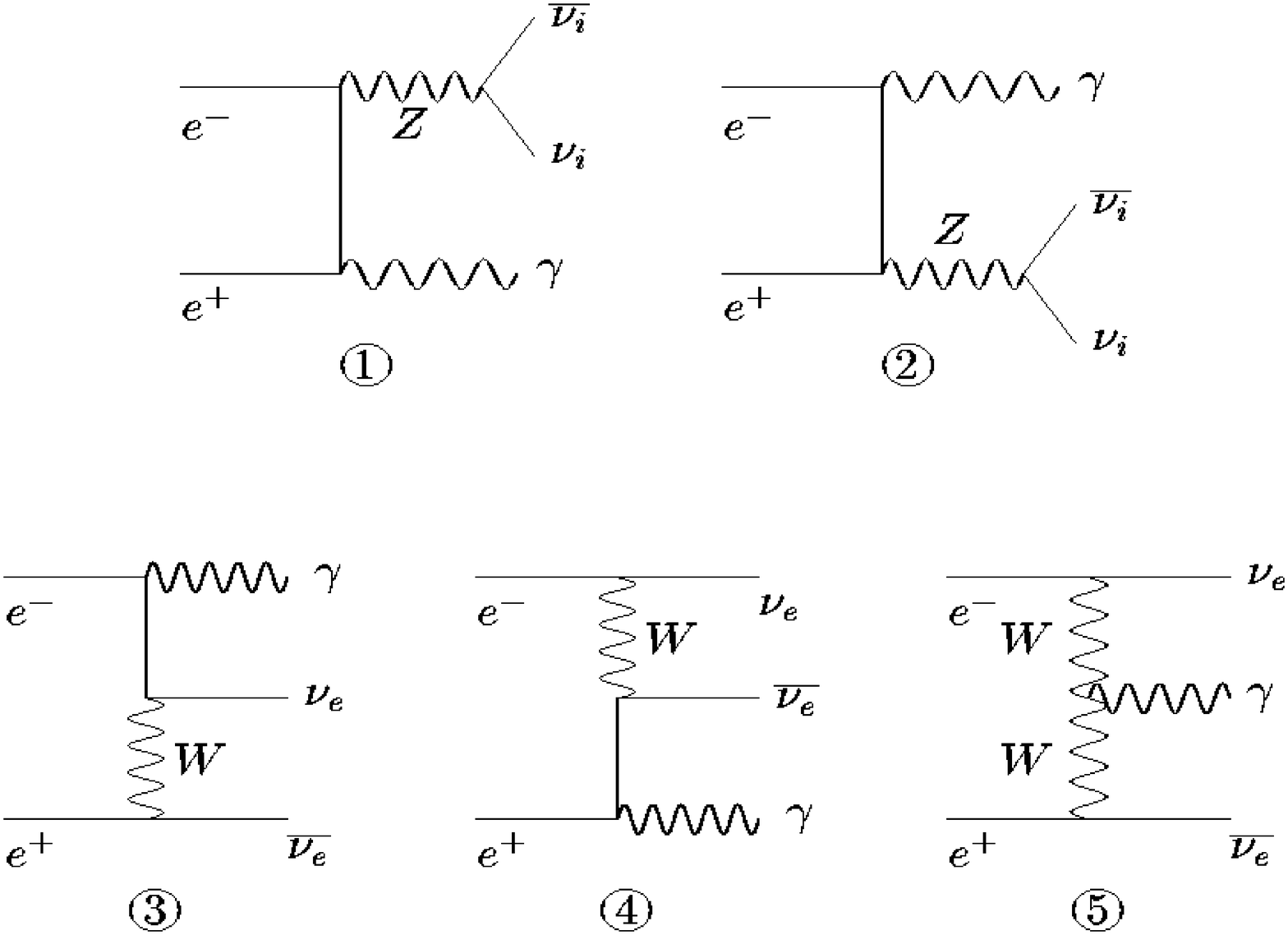,width=65mm}}}
\caption{The process $e^+e^-\to\nu\bar\nu\gamma$.}
\label{fignunug}
\end{figure}

The respective tree level process is given by the Feynman diagrams
in Fig.~\ref{fignunug}.
As described in ref.~\cite{recent1}, the
\KK\ MC with CEEX matrix element is now extended to the neutrino mode.
It is a replacement for the older KORALZ program. 
This new mode of the \KK MC is useful for LEP final data analysis
and for the first steps toward the LC. We note the following properties and
improvements due to this new \KK MC CEEX treatment of the $e^+e^-\to\nu\bar\nu\gamma$ process: (1), the systematic error is now estimated to be
1.3\%  for $\nu_e \bar \nu_e \gamma$ and 0.8\% for 
$\nu_\mu \bar \nu_\mu \gamma $ and $ \nu_\tau\bar \nu_\tau \gamma$; (2),
for observables with two observed photons we estimate the uncertainty to
be about 5\%; (3), these new improved results were obtained thanks to
the inclusion of the non-photonic electroweak corrections
of the {\tt ZFITTER} package
and due to newly constructed, exact,
single and double photon emission amplitudes in the \KK\ MC
for the contribution with the $t$-channel $W$ exchange; and, (4), the virtual corrections for the $W$ exchange are at present introduced in an 
approximated form but
the CEEX exponentiation scheme is the same as in the original \KK\ MC program.

\section{Exact Differential ${\cal O}(\alpha^2)$ Results for Hard Bremsstrahlung in $e^+e^-\to 2f $}

\begin{figure}[htb]
\setlength{\unitlength}{1mm}
\framebox[75mm]{\rule[-21mm]{0mm}{20mm}}
\put(-74,-18){\makebox(0,0)[lb]{\epsfig{file=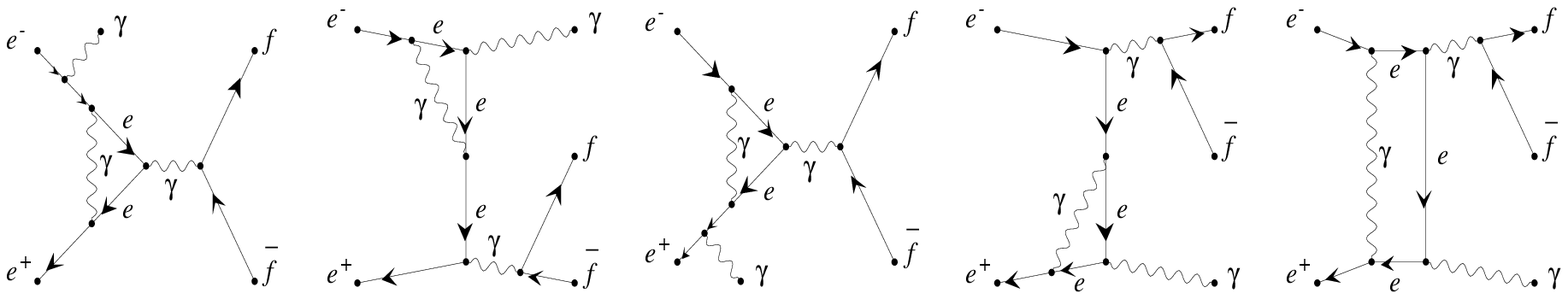,width=73mm}}}
\caption{Representative graphs for the $1\gamma_{real}+1\gamma_{virtual}$ correction in 2f processes.}
\label{figgraphs}
\end{figure}

The respective ${\cal O}(\alpha^2)$ process is illustrated in Fig.~\ref{figgraphs}.
\noindent
In ref.~\cite{recent2}, we have presented 
fully differential results for $2f+1\gamma_{virt}+1\gamma_{real}$ checked against those of Igarishi and Nakazawa in ref.~\cite{in1} and
partly integrated results checked against those of 
Berends, Burgers and Van Neerven in ref.~\cite{berends}. Our results are an
important component for any exact \Order{\alpha^2} 
exponentiated calculation for $e^+e^-\to 2f $.
Similar works were also recently completed by the Karlsruhe group~\cite{hans}
-- a comparison with their results is in progress.

\begin{figure}[htb]
\setlength{\unitlength}{1mm}
\framebox[65mm]{\rule[-21mm]{0mm}{50mm}}
\put(-63,-21){\makebox(0,0)[lb]{\epsfig{file=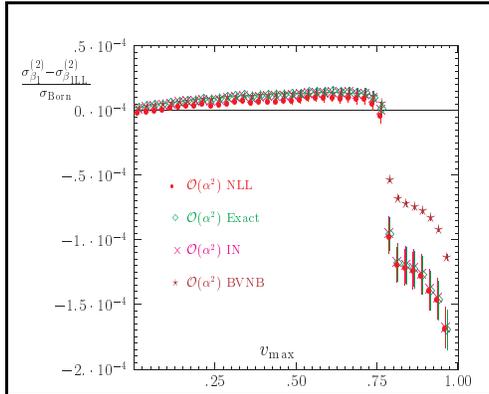,width=60mm}}}
\caption{Exact results for the $1\gamma_{real}+1\gamma_{virtual}$ 
correction in 2f processes.}
\label{figexact}
\end{figure}

We illustrate our results in Fig.~\ref{figexact}.
%
For $v<0.9$ agreement within $0.5 \cdot 10^{-4}$ is reached.

\section{Conclusions}

YFS inspired EEX and CEEX MC schemes are successful examples
of Monte Carlos based directly on the factorization theorem
(albeit for the IR soft case for Abelian QED only). These schemes
work well in practice: KORALZ, BHLUMI, YWSWW3, BHWIDE and \KK MC are
examples. The extension of such schemes (as far as possible) to 
all collinear singularities
would be very desirable and practically important! Work on this
is in progress.

Here, we have illustrated that the \KK MC program is extended to the neutrino channel. Moreover, we have shown that
the missing fully differential $2f+1\gamma_{virt}+1\gamma_{real}$ distributions
for \Order{\alpha^2} CEEX are now available. Applications to
final LEP data analysis and to LC studies are in progress.

The authors thank Profs. G. Altarelli and Wolf-Dieter Schlatter
for the support and kind hospitality of the CERN Theory Division 
and the CERN LEP Collaborations while this work was in progress.
One of the authors (B.F.L.W.) also thanks Profs. S. Bethke and 
L. Stodolsky for the support and kind hospitality of the
Werner-Heisenberg-Institut, Max-Planck-Institut, Munich,
while this work was in progress.


\begin{thebibliography}{9}
\bibitem{ceex1} S. Jadach, B.F.L. Ward Z. W\c{a}s,
Phys. Rev. D {\bf 63} (2001) 113009; Comput. Phys. Commun. {\bf 130}
 (2000) 260; Eur. Phys. J. {\bf C22} (2001) 423; 
Phys. Lett. {\bf B449} (1999) 97, and references therein.
\bibitem{recent1} D. Bardin, S. Jadach, T. Riemann and  Z. Was,
Eur. Phys. J. {\bf C24} (2002) 373, and references therein.
\bibitem{recent2} S. Jadach, M.~Melles, B.F.L. Ward and S. A. Yost,
Phys. Rev. {\bf D65} (2002) 073030, and references therein.
\bibitem{yfs}
D.~R.~Yennie, S.~C.~Frautschi, and H.~Suura, Ann. 
Phys. {\bf 13} (1961) 379;\newline
see also K.~T.~Mahanthappa, Phys.~Rev.~{\bf 126} (1962) 329, 
for a related analysis.
\bibitem{eex}
S. Jadach and B.F.L. Ward, Phys. Rev. {\bf D38} (1988) 2897;{\it ibid.}
{\bf D39} (1989) 1471; {\it ibid.} {\bf D40} (1989) 3582;
S.Jadach, B.F.L. Ward and Z. Was, Comput. Phys. Commun. {\bf 66} (1991) 276;
S.Jadach and B.F.L. Ward, Phys. Lett. {\bf B274} (1992) 470;
S. Jadach {\em et al.}, Comput. Phys. Commun. {\bf 70} (1992) 305;
S.Jadach, B.F.L. Ward and Z. Was, Comput. Phys. Commun. {\bf 79} (1994) 503;
S. Jadach {\em et al.}, Phys. Lett. {\bf B353} (1995) 362; {\it ibid.} 
{\bf B384} (1996) 488; Comput. Phys. Commun. {\bf 102} (1997) 229;
S.Jadach, W. Placzek and B.F.L. Ward, Phys. Lett. {\bf B390} (1997) 298;
Phys. Rev. {\bf D54} (1996) 5434; Phys.Rev. D56 (1997) 6939;
S.Jadach, M. Skrzypek and B.F.L. Ward, Phys. Rev. {\bf D55} (1997) 1206;
See, for example, S. Jadach {\em et al.}, Phys. Lett. {\bf B417} (1998) 326;
Comput. Phys. Commun. {\bf 119} (1999) 272; Phys. Rev. {\bf D61} (2000) 113010;
Phys. Rev. {\bf D65} (2002) 093010; Comput. Phys. Commun. {\bf 140} (2001) 432,
475;
S.Jadach, B.F.L. Ward and Z. Was, Comput. Phys. Commun. {\bf 124} (2000) 233; 
and references therein.
\bibitem{zfitter} D. Bardin {\em et al.}, Comput. Phys. Commun. {\bf 133} 
(2001) 229.
\bibitem{berends} 
F.A.\ Berends, W.L.\ Van Neerven and G.J.H.\ Burgers, Nucl.\ Phys.
{\bf B297} (1988) 429, and references therein.
\bibitem{in1} M. Igarishi and N. Nakazawa, Nucl. Phys. {\bf B288} (1987) 301.
\bibitem{hans} H. Kuhn and G. Rodrigo, hep-ph/0204283; G. Rodrigo {\em et al.},
Eur. Phys. J. C{\bf 22} (2001) 81. 
\end{thebibliography}
\end{document}